\documentclass[intlimits,twoside,a4paper]{article}
\usepackage{graphicx}
\usepackage[T2A]{fontenc}
\usepackage[cp1251]{inputenc}
\usepackage{amsmath,amssymb}
\usepackage{wrapfig}
\usepackage{cmpj}

\issue{2011}{14}{1}{13702}

\doinumber{10.5488/CMP.14.13702}

\title[Band spectrum transformation and temperature dependences \ldots{}]%
{Band spectrum transformation and temperature dependences of thermoelectric power of
 Hg$_{1-x}$R$_{x}$Ba$_{2}$Ca$_{2}$Cu$_{3}$O$_{8+\delta}$ system%
}
\author[O.~Babych \textsl{et al.}]
{O.~Babych\thanks{E-mail: orestbabych@gmail.com}\,, I.~Gabriel,
R.~Lutsiv, M.~Matviyiv, M.~Vasyuk}
\address{ Ivan Franko National University of Lviv,
50 Dragomanov Str., 79005 Lviv, Ukraine}
\authorcopyright{O.~Babych, I.~Gabriel,
R.~Lutsiv, M.~Matviyiv, M.~Vasyuk}
 \date{Received September 28, 2009, in final form May 31, 2010}

\begin{document}

\maketitle

\begin{abstract}
Temperature dependences of thermoelectric power S(T) at
$T>T_{\mathrm{c}}$ of the Hg--\,based high temperature
 superconductors  Hg$_{1-x}$R$_{x}$Ba$_{2}$Ca$_{2}$Cu$_{3}$O$_{8+\delta}$ (R=Re, Pb) have
 been analyzed
 with accounting for strong
 scattering of charge carriers. Transformation of parameters of a narrow conducting band in the region of
 the Fermi level was studied. The existence of correlation between the effective bandwidth
 and the temperature of a superconductive transition $T_{\mathrm{c}}$\, is shown.
\pacs 74.25.Fy, 74.62.Dh, 74.72.Jt
\end{abstract}

\section{Introduction}

When studying high temperature superconductors (HTSC), it is
important to identify correlation between major peculiarities of
the transport process of charge carriers in normal phase
$T>T_{\mathrm{c}}$ and the calculated or model densities of states in
the vicinity of the Fermi level.

According to the band calculations, the Fermi level for YBaCuO,
HgBaCuO and other cuprates is near (on a slope) the narrow peak of
the density of states (DOS) formed by overlapping of the $p$\,--
and $d$--\,bands~\cite{Matth,Novik}. Therefore, the usage of the
narrow band phenomenological model to explain peculiarities in the
behavior of the HTSC materials (see for example~\cite{Moshc}) is
understandable. There are considerable discrepancies concerning
 the role of various atoms and their positions in the
elementary cell in the formation of the conductive band. The data
of band-structure calculations show that the peak in the density
of states exists against the background of a considerably wider
band. However, if the Fermi level is located within this narrow
energy interval, where the value of the density of states is
larger than beyond this interval, then this peak plays a dominat
role in the properties of the normal state and possibly of the
superconductive state as well. The width of this band is of the
order of $k_{\mathrm{B}}T$. Therefore, all its levels can make a
considerable contribution to the transport of electrons.

It should be noted that absolute values, slopes of curves of
temperature dependences of the Hall coefficient $R_{x}$ and
resistivity $\rho$ in particular, vary depending on structure
defects, microcracks, and granularity of the medium. Contribution
of the component that is related to imperfections, to thermoelectric power
is considerably smaller. Experimental data on $S(T)$ obtained by
different authors for samples with equal compositions are close to
each other and good reproducibility of the results is observed.
Therefore, differences in values and dependences with the change of
composition should be explained by specific features of electronic
structure. Moreover, using theoretical expressions, one can
calculate the absolute values of thermoelectric power, whereas
resistivity and the Hall coefficient can be calculated with an
accuracy to a constant because of the lack of necessary data on
the material parameters. Due to the aforementioned features, the
analysis in the present study was based on the temperature
dependences of thermoelectric power.

\begin{figure}[h]
\centerline{\includegraphics[width=0.65\textwidth]{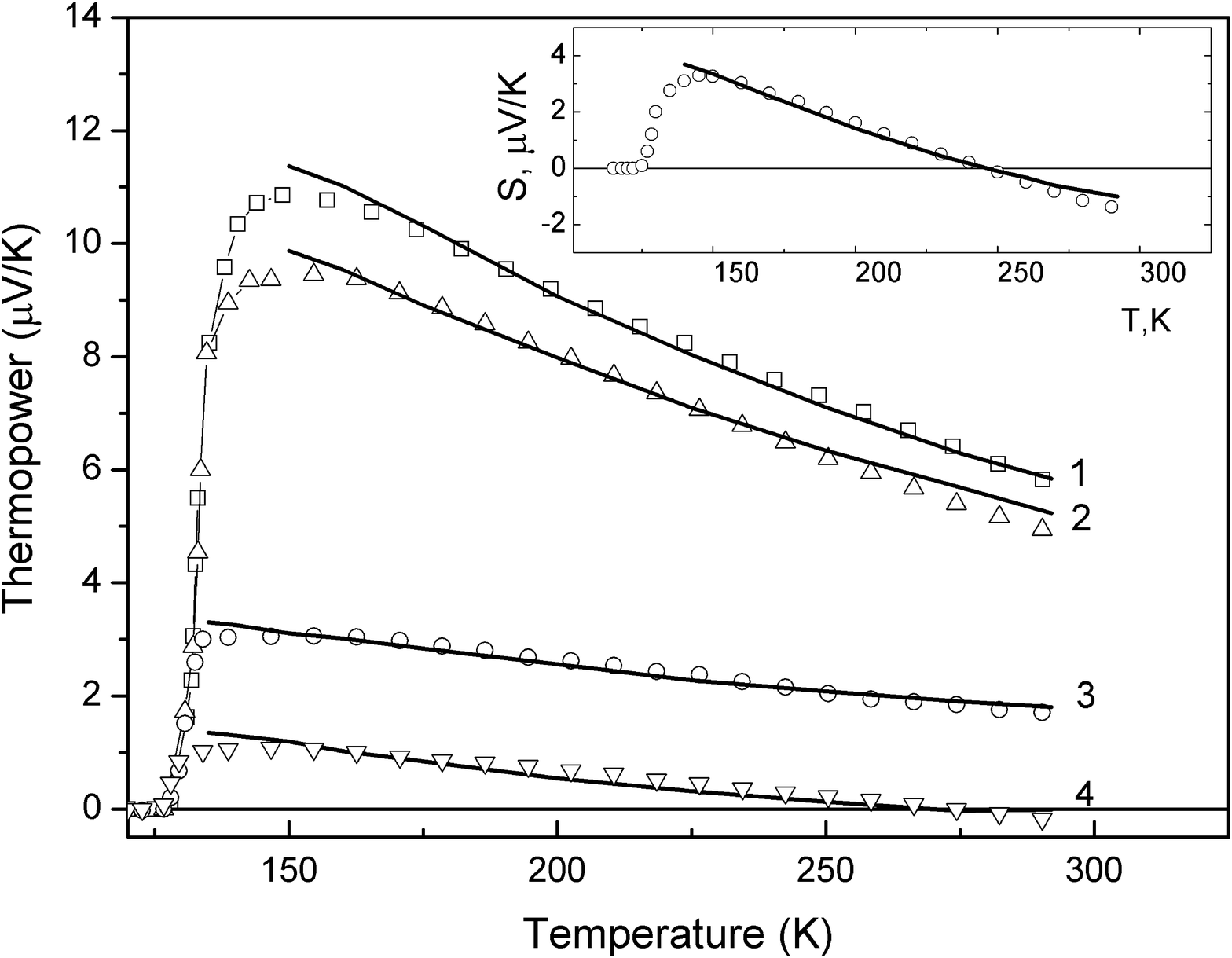}}
\caption{Temperature dependences of thermopower
HgBa$_{2}$Ca$_{2}$Cu$_{3}$O$_{8+\delta}$ annealed under various
conditions~\cite{ChenF}, solid lines~--- calculation
results:\newline 1. $E_{0}$=$-$6.40 meV, $W$=45.9 meV,
$W_{1}/W_{2}$=0.70, $D(E_\mathrm{F})$=21.58 eV$^{-1}$, $F$=0.474,
 $T_\mathrm{c}$=128.7 K;\newline
2. $E_{0}$=$-$4.60 meV, $W$=45.0 meV, $W_{1}/W_{2}$=0.80,
$D(E_\mathrm{F})$=21.96 eV$^{-1}$, $F$=0.488,
 $T_\mathrm{c}$=130.0 K;\newline
3. $E_{0}$=$-$1.18 meV, $W$=43.0 meV, $W_{1}/W_{2}$=0.96,
$D(E_\mathrm{F})$=23.23 eV$^{-1}$, $F$=0.499,
 $T_\mathrm{c}$=133.0 K;\newline
4. $E_{0}$=$-$1.61 meV, $W$=44.1 meV, $W_{1}/W_{2}$=0.87,
$D(E_\mathrm{F})$=22.72 eV$^{-1}$, $F$=0.484,
 $T_\mathrm{c}$=131.6 K;\newline
In insert: experimental (o) and calculation ($-$) data of
thermopower
Hg$_{0.8}$Pb$_{0.2}$Ba$_{2}$Ca$_{2}$Cu$_{3}$O$_{8+\delta}$:\newline
$E_{0}$=$-$4.7 meV, $W$=47.5 meV, $W_{1}/W_{2}$=0.67,
$D(E_\mathrm{F})$=21.50 eV$^{-1}$,  $F$=0.453,
$T_\mathrm{c}$=126.5 K.} \label{fig-1}
\end{figure}
The use of the band densities of states does not always make it possible to explain the
 transport properties of the HTSC. This can be attributed to the
fact that correlation effects are not taken into account in the one
electron approximation of the band theory. Furthermore, there is frequently no
information on electronic structure of compounds with deviations
of their compositions from stoichiometry. That is why
we have used a model presentation of the density of
states in the Lorentz form:
\begin{equation}
\label{ge-def} D(E) = \frac{1}{\pi}\frac{W}{(E-E_{0})^{2}+W^{2}}\, ,
\end{equation}
where $E_{0}$~--- is the distance of the peak of DOS from the
Fermi level, $W$~--- is the width of DOS peak. Thermoelectric
power temperature depedence in the phenomenological model term of
narrow band using reference~\cite{Mott} has been calculated:
\begin{equation}
\label{st-def} S(T) = -\frac{1}{|\re|T} \frac{I_{1}}{I_{0}}\, ,
\end{equation}
where
\begin{equation}
\label{i1-def} I_{1} =
\int_{}^{}\sigma_{E}\left(-\frac{df}{dE}\right)\left(E-E_\mathrm{F}\right)\rd
E,
\end{equation}
\begin{equation}
\label{i0-def} I_{0} =
\int_{}^{}\sigma_{E}\left(-\frac{df}{dE}\right)\rd E \, ,
\end{equation}
where $\sigma_{E}$~--- is the conductivity at $T \to 0$, sensitive
to the fine structure of the density of states near
$E_\mathrm{F}$\,, $f(E - E_\mathrm{F})$~--- the Fermi distribution
function. Resistivity is inversely proportional to:
\begin{equation}
\label{ro-def} \rho(T) \sim 1/I_{0} =
1/\int_{}^{}\sigma_{E}\left(-\frac{df}{dE}\right)\rd E.
\end{equation}
To analyse the results obtained we estimated the degree of band filling by electrons:
\begin{equation}
\label{Ft-def} F(T) = \frac{\int f(E-E_\mathrm{F})D(E)\rd E }{
\int D(E)\rd E }\,.
\end{equation}
When selecting the type of scattering of  carriers, one ought to
note that during the interpretation of analogous dependences of
$S(T)$ in intermetallic systems with intermediate valence and
heavy fermions on the base of the $4f$--\,, $3d$--\,transition
elements, logical results can be obtained by assuming $\sigma _{E}
\sim D^{-1}(E)$ (the Mott model~-- rather weak scattering,
attributed mainly to the $p-d$ transitions)~\cite{Koter,Koter1}.
Values of the mobility of carriers and conductivity for samples of
HTSC are comparatively small~\cite{Chen,Subra}, and thus we
have used the dependence on the basis of the Kubo-Greenwood
formula $\sigma_{E} \sim D^{2}(E)$ (general case of a strong
scattering).

\section{Technical data-out}

Members of the mercury homological series
HgBa$_{2}$Ca$_{n-1}$Cu$_{n}$O$_{2n+n+\delta}$\,, which demonstrate
the highest currently known temperature of transition in a
superconductive state $T_\mathrm{c}$ ($n=3$), of the order of 130
K and 160 K both at atmospheric~\cite{Schill} and increased
pressures~\cite{Chu,Gao}, correspondingly, were selected as objects
for the study. There is a large amount  of published
 experimental data for these superconductive ceramics on
 temperature dependences of thermoelectric power
 during cationic substitutions and anionic doping.

\begin{figure}[htb]
\centerline{\includegraphics[width=0.65\textwidth]{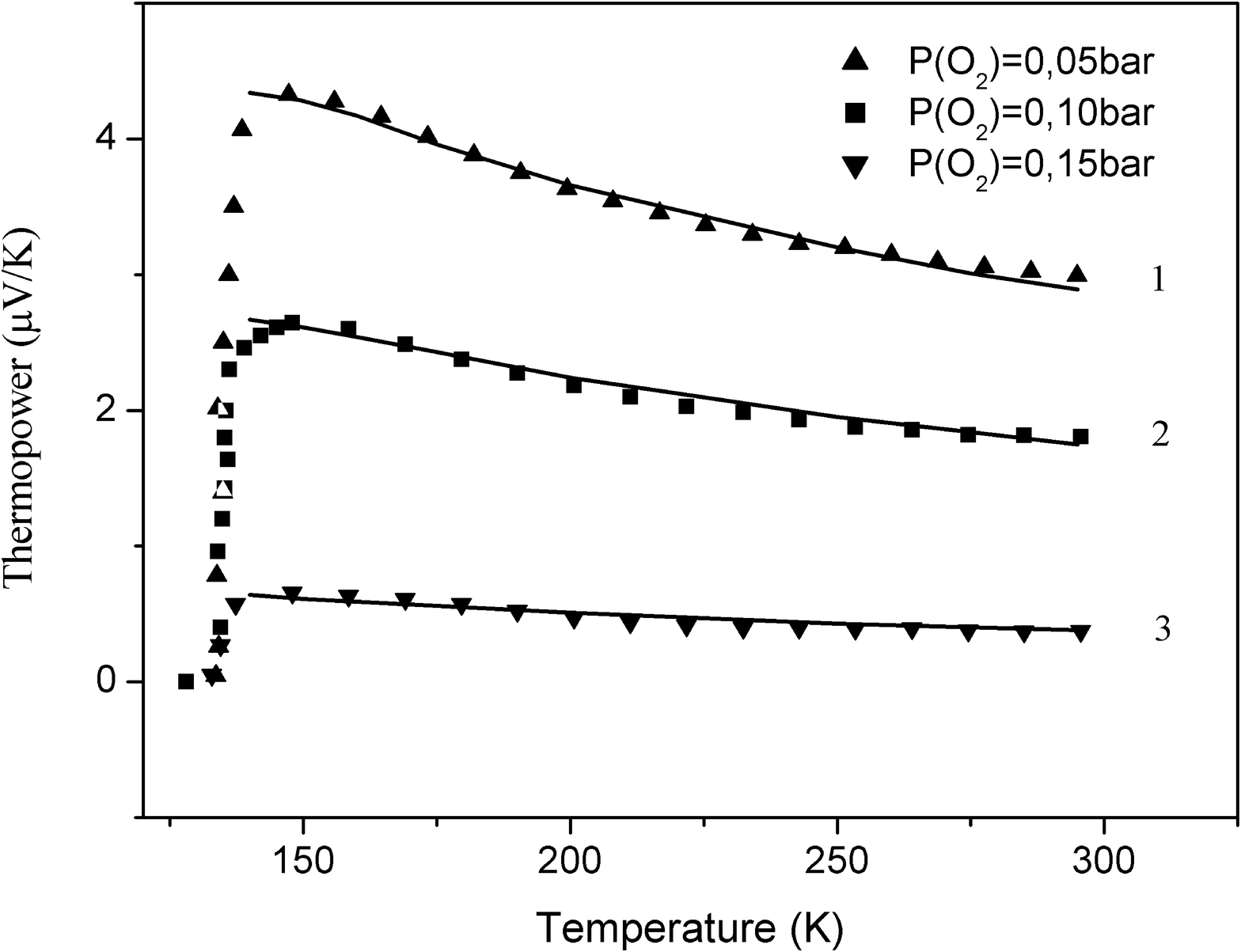}}
\caption{Temperature dependences of thermopower
Hg$_{0.82}$Re$_{0.18}$Ba$_{2}$Ca$_{2}$Cu$_{3}$O$_{8+\delta}$
 annealed under various conditions~\cite{Passos}, solid lines~--- calculation results:\newline
1. $E_{0}$=$-$1.08 meV, $W$=43.3 meV, $W_{1}/W_{2}$=0.99,
$D(E_\mathrm{F})$=23.06 eV$^{-1}$, $F$=0.506,
 $T_\mathrm{c}$=132.6 K;\newline
2. $E_{0}$=$-$0.63 meV, $W$=42.8 meV, $W_{1}/W_{2}$=1.00,
$D(E_\mathrm{F})$=23.36 eV$^{-1}$, $F$=0.504,
 $T_\mathrm{c}$=133.2 K;\newline
3. $E_{0}$=$-$0.20 meV, $W$=43.2 meV, $W_{1}/W_{2}$=0.99,
$D(E_\mathrm{F})$=23.16 eV$^{-1}$, $F$=0.500,
 $T_\mathrm{c}$=132.7 K.}
\label{fig-2}
\end{figure}

The experimental data for HgBa$_{2}$Ca$_{2}$Cu$_{3}$O$_{8+\delta}$
(figure~\ref{fig-1}) and
Hg$_{0.82}$Re$_{0.18}$Ba$_{2}$Ca$_{2}$Cu$_{3}$O$_{8+\delta}$
(figure~\ref{fig-2}) are available from~\cite{ChenF,Passos}.
Temperature dependences of thermoelectric power features with
clearly visible maximum at temperatures above the superconductive
transition, are presented as a section of practically linear
fall-off  with temperature increasing up to 290 K. Authors of this
paper synthesized the samples and measured $\rho (T)$,
$T_\mathrm{c}$ and $S(T)$ when Hg is substituted for Pb (see
figure~\ref{fig-1}, insert). Introduction of Re, Pb and some other
elements results in an improvement of chemical stability without
considerable losses of initial value for $T_\mathrm{c}$\,. Values
of the thermoelectric power decrease up to possible sign inversion
for certain compositions with an increase of oxygen content both in
pure and doped samples.

\section{Results and discussion}

Analyzing the temperature dependences of  thermoelectric power in
the model of a narrow band and reaching a qualitative matching
between the calculated and experimental data, it is possible to
estimate the band spectrum parameters and trace their
transformation with variations of composition of the samples.

Representation of the peak of the density of states in the form of a
symmetric Lorentzian~(\ref{ge-def}) does not yield satisfactory
results. Good agreement of the calculation with experimental
dependences at $T>T_\mathrm{c}$\,, in particular inversion of the
thermoelectric conductivity sign, can be obtained using a
 asymmetric peak of the density of states, which is defined by a
position of the $E_{0}$ maximum relatively to the Fermi level and
the two half band widths $W_{1}$ at $E<E_{0}$ and $W_{2}$ at
$E>E_{0}$. The average value of $W=(W_{1}+W_{2})/2$ is presented
in the paper. Let us note the following point: if  the band
filling degree $F>1/2$, then symmetric Lorentz distribution is
below $E_\mathrm{F}$ and $S>0$; if $F<1/2$, it is higher than
$E_\mathrm{F}$ and $S<0$, and if $F=1/2$~--- $S=0$. In the case of
an asymmetric band, positive values of $S$ can be  also observed at
$F \leqslant 1/2$.
\begin{figure}[htb]
\centerline{\includegraphics[width=0.57 \textwidth]{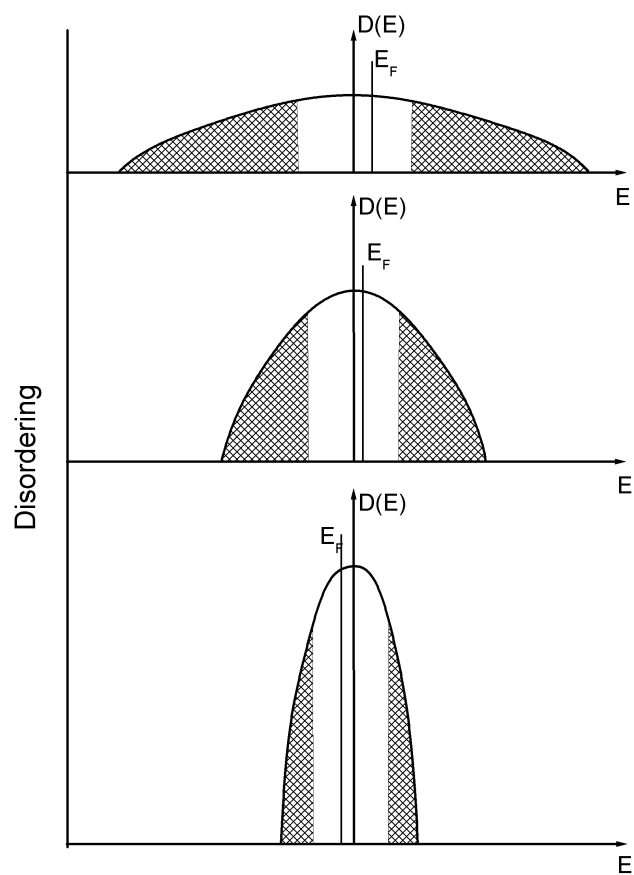}}
\caption{Band broadening, increase of the part of localized states
at the band edges (shaded areas), fall of the density of states at
the Fermi level $D(E_\mathrm{F})$ as a function of structure
disordering.} \label{fig-3}
\end{figure}

The results of the calculations using the approach described above
for HgBa$_{2}$Ca$_{2}$Cu$_{3}$O$_{8+\delta}$ in the region of
optimal oxygen doping are shown in figure~\ref{fig-1}. Here, the
degree of band filling  decreases in the whole range when the
oxygen index $\delta$ grows, which corresponds to the acceptor
type of additional ions of oxygen introduced in the system.  When
the bandwidth increases in the vicinity of the optimal doping
region (curves 2 and 4 in figure~\ref{fig-1}), the density of
states at the Fermi level $D(E_\mathrm{F})$ decreases, which
correlates with a decrease of $T_\mathrm{c}$\,. This can be
related to the lattice disordering (increase of structural
defects) when oxygen content deviates from the optimal one (curve
3) both downwards (curve 2) and upwards (curve 4). This is
confirmed by an increase of the band asymmetry ($W_{1}/W_{2}$) for
compositions 2, 4 as compared with composition 3. Consequently,
the largest $T_\mathrm{c}$ (maximal ordering degree) corresponds
to sample 3.

An analogous type of the band spectrum transformation was obtained
during the analysis of temperature dependences of thermoelectric power
of the
Hg$_{0.82}$Re$_{0.18}$Ba$_{2}$Ca$_{2}$Cu$_{3}$O$_{8+\delta}$
system with variation of the oxygen index $\delta$
(figure~\ref{fig-2}). Here, correlation between the conducting
band width and correspondingly density of states on the
$E_\mathrm{F}$ and the temperature of transition in
superconductive state is also observed, i.e., larger values of
$T_\mathrm{c}$ for composition 2 as compared to the 1 and~3.

The relationships obtained correspond to the Anderson model: band
broadening, increase of the part of localized states at the band
edges, the drop of the density of states at the Fermi level
$D(E_\mathrm{F})$ as a function of disordering structure
(figure~\ref{fig-3}), which corresponds to a type of
transformation of the band spectrum and temperatures of
superconductive transition obtained from calculation.

Comparing the dependences $S(T)$ for pure Hg--1223 samples and
those optimally doped with oxygen (figure~\ref{fig-1}) with the
experimental obtained data for
Hg$_{0.8}$Re$_{0.2}$Ba$_{2}$Ca$_{2}$Cu$_{3}$O$_{8+\delta}$ (insert
in the figure~\ref{fig-1}), it is found that the inversion is only
observed for the (Hg,Pb)--1223 samples for one level
$\alpha_\mathrm{max} \cong 3$~$\mu$V/K with the increase of
temperature, which points to additional introduction of holes due
to cationic substitution. This is confirmed by the calculation
results: with such substitution of mercury by lead, there takes
place a decrease of the degree of band filling by electrons, a
growth of its symmetry and width, and correspondingly a decay of
the $D(E_\mathrm{F})$, which corresponds to the change of
$T_\mathrm{c}$ (figure~\ref{fig-1}).

\begin{figure}[htb]
\centerline{\includegraphics[width=0.65\textwidth]{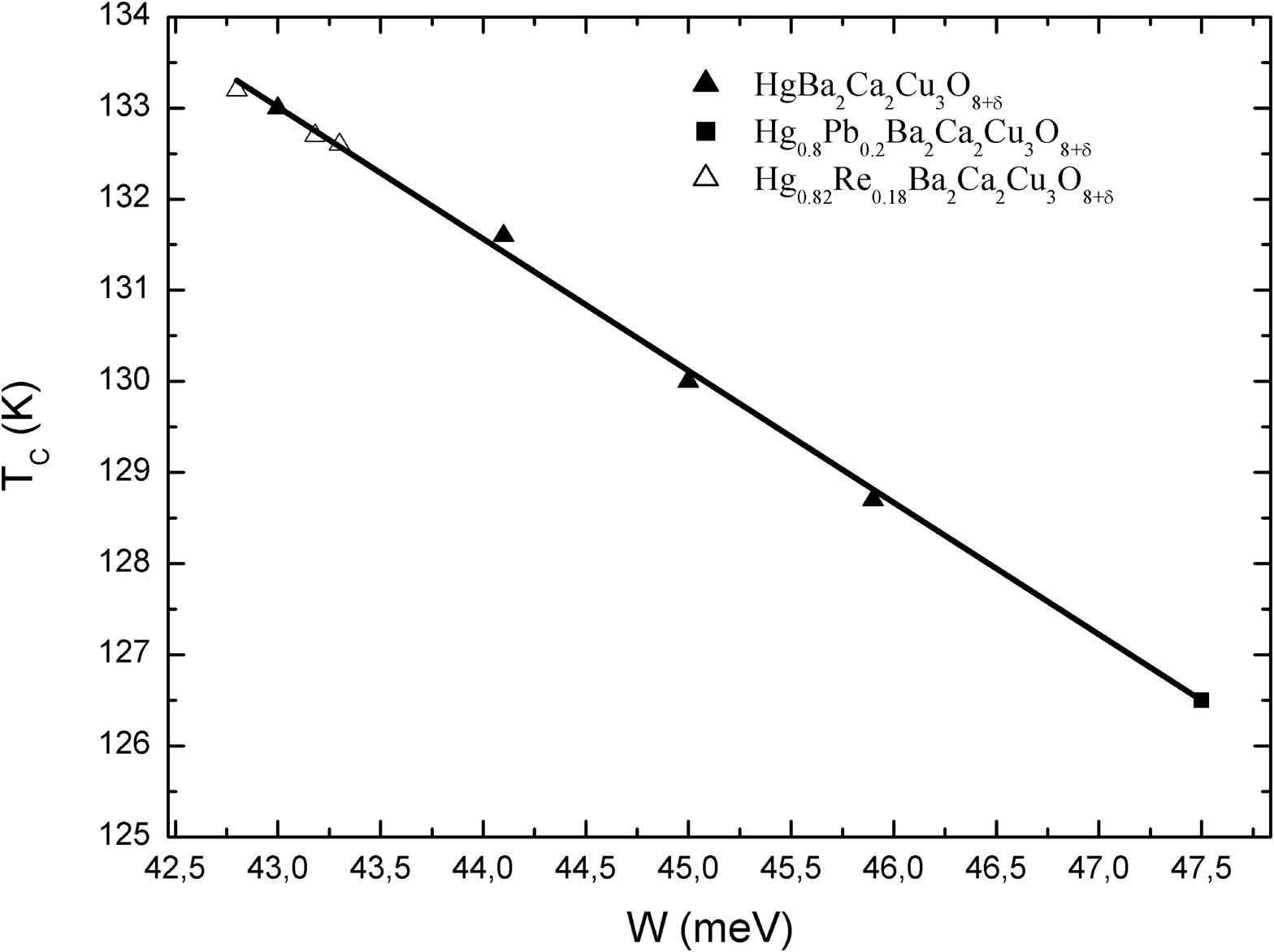}}
\caption{Correlative dependence between critical temperature and
effective conducting bandwidth $T_\mathrm{c}(W)$ in the system of
Hg$_{1-x}$R$_{x}$Ba$_{2}$Ca$_{2}$Cu$_{3}$O$_{8+\delta}$ (R=Re,
Pb).} \label{fig-4}
\end{figure}

The results of calculations enabled us to assess an impact of various
cationic and anionic substitutions in Hg--\,containing HTSC not
only on the band spectrum parameters, but also on critical
temperature, as well as to trace the relationship between them.
Figure~\ref{fig-4}  graphically depicts the correlation between
$T_\mathrm{c}$ and the bandwidth $W$. Reduction of the density of
states value at the Fermi level due to the band broadening can be
caused by a decrease of the $T_\mathrm{c}$\,.

\section{Conclusion}

Using a representation of the peak of the density of states in the
region of the Fermi level in the form of an asymmetric Lorentz
distribution, good agreement of the calculated temperature
dependences of the thermoelectric power with experimental data at
$T>T_\mathrm{c}$ has been obtained. The type of the behaviour of
thermoelectric power of the Hg--\,containing high temperature
superconductors being studied, with cationic substitution and
anionic doping, is defined by such parameters of the narrow
conduction band in the region of the Fermi level as its width $W$,
population by electrons $F$ and values of the density of states
$D(E_\mathrm{F})$. Existence of correlation between the effective
band width and the temperature of superconductive transition has
been shown during establishing interrelation between properties of
the normal state and superconductive characteristics.


\ukrainianpart

\title{Трансформація зонного спектру та температурні залежності коефіцієнта термоелектрорушійної сили в системі
 Hg$_{1-x}$R$_{x}$Ba$_{2}$Ca$_{2}$Cu$_{3}$O$_{8+\delta}$ }
\author{О. Бабич, І. Габрієль, Р. Луців, М. Матвіїв, М. Васюк}
\address{Львівський національний університет імені Івана Франка, вул. Драгоманова, 50, Львів, Україна
}
%
%
%

\makeukrtitle

\begin{abstract}
\tolerance=3000%
З врахуванням сильного розсіювання носіїв заряду проведено аналіз температурних залежностей коефіцієнта термоелектрорушійної сили S(T) при Т>T$_{\rm c}$ ртутьвмісних високотемпературних над\-провідників (ВТНП) Hg$_{1-x}$R$_{x}$Ba$_{2}$Ca$_{2}$Cu$_{3}$O$_{8+\delta}$ (R=Re, Pb). Розглянута трансформація параметрів вузької провідної зони в ділянці рівня Фермі. Показано  існування кореляції між ефективною шириною зони та температурою надпровідного переходу $T_{\rm c}$.
\keywords  високотемпературна надпровідність, температура надпровідного переходу, коефіцієнт термоелектрорушійної сили, вузька провідна зона, пік густини станів, рівень Фермі.
\end{abstract}

\end{document}